\documentstyle[12pt,epsf]{article}
\newcommand{\be}{\begin{equation}}
\newcommand{\ee}{\end{equation}}
\newcommand{\bea}{\begin{eqnarray}}
\newcommand{\eea}{\end{eqnarray}}
\newcommand{\bt}{\begin{tabular}}
\newcommand{\et}{\end{tabular}}
\newcommand{\ba}{\begin{array}}
\newcommand{\ea}{\end{array}}

\newcommand{\bvec}{\vec}

\def\gr{\hbox{$\nabla$}}
\def\di{\hbox{$\nabla \, {\cdot} \,$}}
\def\ro{\hbox{$\nabla \, {\times} \,$}}
\def\la{\hbox{$\nabla^2$}}

\newcommand{\dt}{\partial t}

\newcommand{\vb}{\bvec{v}_B}
\newcommand{\vs}{\bvec{v}_S}
\newcommand{\x}{\bvec{x}}
\newcommand{\s}{\bvec{s}}

\begin{document}
\setcounter{page}{0}
\thispagestyle{empty}
\baselineskip=20pt

\hfill{
\begin{tabular}{l}
DSF$-$98/15 \\
INFN$-$NA$-$IV$-$98/15 \\
\end{tabular}}

\bigskip\bigskip

\begin{center}
\begin{Large}
{\bf On the role of Spin in Quantum Mechanics}
\end{Large}
\end{center}

\vspace{0.5truecm}

\begin{center}
{\large
Salvatore Esposito
\footnote{e-mail: sesposito@na.infn.it}}
\end{center}

\normalsize
\begin{center}
{\it
\noindent
Dipartimento di Scienze Fisiche, Universit\`{a} di Napoli ``Federico II''\\ and
\\ Istituto Nazionale di Fisica Nucleare, Sezione di Napoli\\ Mostra
d'Oltremare Pad. 20, I-80125 Napoli Italy }
\end{center}

\vspace{1truecm}

\begin{abstract}
\noindent
From the invariance properties of the Schr\"{o}dinger equation and the isotropy
of space we show that a generic (non-relativistic) quantum system is
endowed with an ``external'' motion, which can be interpreted as the motion
of the centre of mass, and an ``internal'' one, whose presence disappears
in the classical limit. The latter is caused by the spin of the particle,
whatever is its actual value (different from zero). The quantum potential
in the Schr\"{o}dinger equation, which is responsible of the quantum effects of
the system, is then completely determined from the properties of the
internal motion, and its ``unusual'' properties have a simple and physical
explanation in the present context. From the impossibility to fix the
initial conditions relevant for the internal motion follows, finally, the
need of a probabilistic interpretation of quantum mechanics.
\end{abstract}

\vspace{0.5truecm}

\newpage

\section{Introduction}

A quantum elementary system is described by a complex wave field
$\psi (\x , t)$ which, in the non-relativistic limit, satisfies the
Schr\"{o}dinger equation
\be
i \, \frac{\partial \psi}{\dt} \; = \; \mathrm{H} \, \psi
\label{11}
\ee
where H is the hamiltonian operator
\be
\mathrm{H} \; = \; - \, \frac{\la}{2 m} \; + \; U
\label{12}
\ee
$U$ being the (external) potential experienced by the system of mass
$m$ (in this paper we use natural units, in which $\hbar \, = \, c \,
= \, 1$). The complex equation (\ref{11}) can be equivalently written
as two real equations for the modulus $R$ and the phase $S$ of the
function $\psi$:
\bea
\frac{\partial S}{\dt} \; + \; \frac{(\gr S)^2}{2 m} \; - \;
\frac{1}{2 m} \frac{\la R}{R} \; + \; U & = & 0
\label{13} \\
\frac{\partial R^2}{\dt} \; + \; \di \left( \frac{1}{m} \, R^2 \, \gr
S \right) & = & 0
\label{14}
\eea
The last equation is usually referred to as the continuity equation
for the probability density $R^2 \, = \, |\psi|^2$. Instead, eq.
(\ref{13}) has the form of an Hamilton-Jacobi equation for the
characteristic function $S$ of a system described by an effective
potential
\be
V \; = \; U \; + \; Q \; = \; U \; - \; \frac{1}{2 m} \frac{\la R}{R}
\label{15}
\ee
The term $Q$ is called the ``quantum potential''; it is the only
non-classical term (i.e. proportional to the Planck constant)
entering in the set of equations (\ref{13}),(\ref{14}).

Recently, an important paper appeared \cite{rs} in which Recami and Salesi
give a straightforward interpretation of the quantum potential term.
Starting from the Gordon decomposition (in the non-relativistic limit) of
the Dirac current for a spin 1/2 particle, they have shown that the quantum
potential term is strictly related to the spin of the particle and it can
be derived from the kinetic energy associated with the internal
zitterbewegung motion. This result puts new light on the whole
interpretation of quantum mechanics.

However, the mentioned paper is too much related to spin 1/2
particles, so that more general conclusions cannot, in principle, be
drawn.

In the present paper we generalize the Recami and Salesi result by starting
only from the invariance properties of the field equations
(\ref{13}),(\ref{14}). The interpretation of the quantum potential as a
kinetic energy for an internal motion is here achieved for a system of
arbitrary spin. In section 2 a simple formalism is introduced in which the
analysis of field equations is particularly useful for our purposes. This
analysis is carried in section 3, in which the invariance properties of the
equations of motion are studied; these naturally lead, in section 4, to the
identification of an ``internal motion'' of the system, giving origin to
the quantum potential. This identification is made possible by the presence
of the spin, even if it is strictly related to its direction but not to its
actual value (this property allows the generalization of Recami and Salesi
result). In section 5 some ``unusual'' properties of the quantum potential
are analyzed and it is shown that in the present interpretation they
acquire a very simple meaning. Moreover, even if our result does not deal
with the statistical \cite{stat} or causal \cite{holland} interpretation of
quantum mechanics, it is nevertheless particularly useful in the context of
the latter one. Then, in section 6 we reformulate the De Broglie - Bohm
causal interpretation of quantum mechanics in the light of the obtained
results. In particular, the consequences of the insurgence of an ``internal
motion'' for a probabilistic interpretation are pointed out. Finally, in
section 7 there are our conclusions and remarks.

\section{Definition of the variables}

Let us write the wave field in the polar representation
\be
\psi (\x , t) \; = \; R (\x , t) \, e^{i \, S(\x , t)}
\label{21}
\ee
with $R,S$ two real functions. It is useful to introduce also
\be
\rho (\x , t) \; = \; R^2 (\x , t) \; = \; \psi^{\ast} \, \psi
\label{22}
\ee
Taking the logarithmic gradient of (\ref{21}) one obtains
\be
\frac{1}{\psi} \, \gr \psi \; = \; \frac{1}{2} \, \frac{1}{\rho} \,
\gr \rho \; + \; i \, \gr S
\label{23}
\ee
so that
\bea
\rho \, \gr S & = & Im( \psi^{\ast} \, \gr \psi )  \label{24}\\
\gr \rho & = & 2 \, Re(\psi^{\ast} \, \gr \psi )  \label{25}
\eea
Let us now define the following two fields:
\bea
\vb & = & \frac{1}{m} \, \gr S \label{26} \\
\vs & = & \frac{1}{2m} \, \frac{1}{\rho} \, \gr \rho  \label{27}
\eea
Both are adimensional quantities; by definition they are irrotational
fields:
\bea
\ro \vb & = & 0  \label{28} \\
\ro \vs & = & 0 \label{29}
\eea
Their mutual scalar and vector multiplication can be expressed in terms of
the wave field in the following way, respectively
\bea
\vb {\cdot} \vs & = & \frac{1}{2 m^2} \, Im \left\{ \left( \frac{1}{\psi}
\, \gr \psi \right)^2 \right\} \label{212} \\
\vb {\times} \vs & = & \frac{1}{2 m^2} \, \frac{\gr \psi^{\ast} \,
{\times} \, \gr \psi}{\psi^{\ast} \psi} \label{213}
\eea
Note that, even if $S(\x , t)$ is a multi-valued function (it is a
phase), $\vb (\x , t)$ is a single-valued one. However, in nodal
points, where $\psi$ (and then $R$) vanishes, $S$ is undefined, and so
$\vb$. In this case also equation (\ref{28}) is no longer valid; a
general expression can be found in \cite{taka}. \\

\section{Dynamical properties}

With the formalism settled in the previous section, we are now able to
reconsider the Schr\"{o}dinger equation (\ref{11}) by starting, as in
\cite{rs}, from the lagrangian for a non relativistic scalar particle:
\begin{equation}\label{31}
  {\cal L} \; = \; \frac{i}{2} \, \left( \psi^\ast \,
  \frac{\partial}{\partial t} \, \psi \; - \; \left(
  \frac{\partial}{\partial t} \, \psi^\ast \right) \psi \right)
  \; - \; \frac{1}{2 m} \left( \gr \psi^\ast \right) {\cdot}
  \left( \gr \psi \right) \; - \; U \, \psi^\ast \psi ~~~~.
\end{equation}
In terms of the fields $\vb$, $\vs$ we then have
\be \label{32}
{\cal L} \; = \; - \, \rho \, \left( \frac{\partial S}{\dt} \; + \;
\frac{1}{2} \, m \, \vb^2 \; + \; \frac{1}{2} \, m \, \vs^2 \; + \; U
\right) \nonumber
\ee
and from this, the equations of motion (\ref{13}), (\ref{14}) follow:
\bea
\frac{\partial S}{\dt} & = & - \, {\cal H} \label{33} \\
\frac{\partial \rho}{\dt} & = & - \, \di \bvec{J} \label{34}
\eea
with
\bea
{\cal H} & = & \frac{1}{2} \, m \, \vb^2 \; + \; \left(
\frac{1}{2} \, m \,\vs^2 \; - \; \frac{1}{4 m} \, \frac{\la \rho}{\rho}
\right) \, + \; U  \label{35} \\
\bvec{J} & = & \rho \, \vb \label{36}
\eea
The equations of motion are invariant under the transformations
\bea
R( \x , t ) & \longrightarrow & N \, R(\x,t) \\
S( \x , t ) & \longrightarrow & S( \x , t) \; + \; a  \label{312}
\eea
($N$ and $a$ being two constants). The first expresses the invariance
under normalization change, while the second
one correspond to a global phase transformation for
the wave function
\be
\psi ( \x , t) \; \longrightarrow \; e^{i \, a} \, \psi (\x , t)
\label{314}
\ee
Both have been extensively studied in the literature. Instead, let us
concentrate on another transformation under which eq. (\ref{34}) is left
invariant:
\be
\bvec{J} ( \x , t) \; \longrightarrow \; \bvec{J} ( \x , t) \; + \; \ro
\bvec{b} (\x , t)  \label{313}
\ee
($\bvec{b}(\x,t)$ being an arbitrary vector field).
Firstly, we observe that the field $\bvec{b}$ can be written, in
general, as
\be
\bvec{b} \; = \; \psi^{\ast} \, \bvec{c} \, \psi
\label{315}
\ee
where $\bvec{c}$ is an arbitrary vector operator. However, here we
will consider the most simple case in which $\bvec{c}$ is a constant
multiplicative vector operator (independent on position), $\bvec{b} \;
= \; \rho \, \bvec{c}$, so that
\be
\rho \, \vb \; + \; \ro \bvec{b} \; = \; \rho \, \left( \vb \; + \;
\left( \frac{1}{\rho} \, \gr \rho \right) \, {\times} \, \bvec{c}
\right) \; = \; \rho \, \left( \vb \; + \; 2 m \, \vs \, {\times} \,
\bvec{c} \right)
\label{316}
\ee
Denoting, for simplicity, $2 m \, \bvec{c} \, = \, \s$, we have then found
that equation (\ref{34}) is satisfied by taking
\be
\bvec{J} \; = \; \rho \, \left( \vb \; + \; \vs \, {\times} \, \s
\right)
\label{317}
\ee
$\s$ being an arbitrary constant vector. So, the current density
vector $\bvec{J}$ can always be written as
\be
\bvec{J} \; = \; \rho \, \bvec{v}
\label{318}
\ee
but in general the velocity field $\bvec{v}$ has two terms:
\be
\bvec{v} \; = \; \vb \; + \; \vs \, {\times} \, \s
\label{319}
\ee

\section{Identification of the vector $\s$}

The vector $\s$ entering in the expression for the current density is
completely arbitrary, but it is suitable of a physical impressive
interpretation in relation to the equation (\ref{32}) for the lagrangian.
In fact, let us postulate that $\bvec{v}$ in (\ref{319}) is the absolute
velocity of the system, and that (as in \cite{rs}) the sum of the second
and third term in (\ref{32}) is nothing that the total kinetic energy of
the system:
\be
 \frac{1}{2} \, m \, \vb^2 \; + \; \frac{1}{2} \, m \, \vs^2 \; = \;
 \frac{1}{2} \, m \, \bvec{v}^2
\label{320}
\ee
With this choice, the vector $\s$ is no longer arbitrary since, by
\be
\frac{1}{2} \, m \, \bvec{v}^2 \; = \; \frac{1}{2} \, m \, \left( \vb^2 \; +
\; \vs^2 \, \s^2 \; - \; \left( \vs \, {\cdot} \s \right)^2 \; + \; 2 \,
\vb \, {\cdot} \, \left( \vs \, {\times} \, \s \right) \right)
\label{321}
\ee
the identification (\ref{320}) is possible only if the following
conditions are satisfied:
\bea
\s^2 & = & 1  \label{322} \\
\vs \, {\cdot} \, \s & = & 0 \label{323} \\
\vb \, {\cdot} \, \left( \vs \, {\times} \, \s \right) & = & 0 \label{324}
\eea
These constraints show that $\s$ has to be a unitary vector orthogonal
to $\vs$, lying in the same plane of $\vb$, $\vs$ ; then it individuates
only a direction (the versus remains unspecified). Moreover, we note
that the picture emerging from the identification (\ref{320}) leads us
to consider $\s$ as a property associated to the particle more than to
the motion of it (it is, in some sense, on the same ground as the mass
$m$, for example). But, from the isotropy of the space, the only
privileged direction that can be associated with a particle is that of
its spin. So, we see that the identification (\ref{320}) has a
physical meaning only if we identify the vector $\s$ with the spin
vector of the particle.\\
Furthermore, the condition (\ref{322}) allows us to consider valid
this result whatever is the actual value of the spin of the particle;
this feature is responsible of the generalization of Recami and Salesi
result \footnote{In ref. \cite{rs} the authors obtained, from the
identification (\ref{320}), also the value of the modulus of the
vector $\s$, consistent with a spin 1/2 particle. However, this was
achieved because they started from the expression for the current
$\bvec{J}$ which is peculiar (and valid) for spin 1/2 particles.
Consistently, only this result they was able to obtain.}. \\
Instead, the condition (\ref{324}) expresses the fact that
\be
\bvec{v} \; = \; \vb \; + \; \vs \, {\times} \, \s \; = \; \bvec{v}_{||}
\; + \; \bvec{v}_{\perp}
\label{325}
\ee
is the decomposition of the absolute velocity $\bvec{v}$ along $\vb$ and
in the direction orthogonal to it. Remembering, then, that the angular
momentum of a particle is always orthogonal to the velocity of the
particle itself, we see that $\bvec{v}_{\perp}$ can be interpreted as
the velocity associated to the spin angular momentum (note that from
(\ref{322}), (\ref{323}) we have $|\bvec{v}_{\perp}| \; = \; | \vs
|$). \\
The emerging picture is then the following: the motion of a quantum
particle can be decomposed into an ``external'' motion, whose velocity
field is $\vb$, and an ``internal'' one, driven by $\vs \, {\times} \,
\s$. The spin of the particle, whose direction is given by $\s$, is the
angular momentum associated to the internal motion. The kinetic energy of
the system is simply given by the sum of the kinetic energies associated to
the two (partial) motions.

As a corollary, we have also that the Schr\"{o}dinger equation contains the
spin of the given particle, but in a subtle way. In fact, it enters only in
the description of the internal motion, but not in that of the external
one; for the latter, as it is well known, we need multi-component wave
functions (for example, a non-relativistic Pauli system is fully described
by a two-component spinor) \footnote{In this paper, for simplicity, we have
used a single-component wave function; for the generalization to spinors
see \cite{holland}.}.

\section{On the properties of the ``quantum potential''}

The quantum description of a given system can be carried out by the
Schr\"{o}dinger equation (\ref{11}) or equivalently by the set of two
coupled equations (\ref{13}), (\ref{14}). In the latter case, the pure
quantum (non classical) effects are described by the quantum potential
$Q$. Even if the two formulations are completely equivalent, the
description in terms of an effective potential suffers for some
``unusual'' properties of the quantum potential itself (for a detailed
discussion, see for example ref. \cite{holland}); nevertheless,
quantum phenomena are indeed ``unusual'', and this question is not
very remarkable. Anyway, here we want to comment on the following
three properties:
\begin{itemize}
\item  Classically-free motion is not, in general, a free motion in
quantum mechanics, due to the presence of the quantum potential;
\item  A classical (external) potential is a given function of the
coordinates. Instead, the quantum potential is derived from the total
quantum state of the system, so that an infinite number of different
forms (associated with the same physical problem) can be generated by
linearly superposing solutions of the Schr\"{o}dinger equation;
\item  The quantum potential is not altered by a change of the
normalization factor of t he wave field $\psi$. The boundary condition
$\psi ( x \rightarrow  \infty ) \rightarrow 0$ does not
necessarily imply that the quantum potential is ineffective at the
infinity.
\end{itemize}
In the present interpretation, the pure quantum term is not properly a
potential term; in fact, from (\ref{35}) we have that
\begin{equation}\label{37}
  Q \; = \; - \frac{1}{2} \, m \,\vs^2 \; - \; \frac{1}{2} \, \di \vs
\end{equation}
and so it is completely determined by the velocity field $\vs$. This point
is a fundamental one: while the potential is an external property
describing the physical system, a velocity derived energy is a property of
the motion (in some sense, an intrinsic property) univocally determined
once the external conditions (i.e. the potential and the initial
conditions) are given. Hence, properties that can appear ``unusual'' for a
potential term can nevertheless be peculiar for a kinetic term:
\begin{itemize}
\item To the classical free motion corresponds the external quantum free
motion (ruled by $\vb$). Instead, the internal motion has no classical
analog (being a pure quantum effect) and comes out from the presence
of the spin;
\item The internal
motion of the system has to depend on the total state of the system, being
properly an intrinsic property. As well as the classical motion is
univocally determined once the potential and the initial conditions are
given, the complete quantum motion is univocally determined once both the
external and the internal motion are given, i.e. once the potential, the
initial conditions and the quantum state of the system are given;
\item The invariance under normalization change is a necessary
condition for the present interpretation, since the internal motion
(ruled by $\vs$, which is invariant) cannot depend on the choice of the
normalization constant, that does not change the state of the system.
Furthermore, the internal motion, being not directly related to
external agents, can in general be non trivial also at the infinity.
\end{itemize}

\section{Reformulation of the causal interpretation of quantum
mechanics}

The results obtained here does not involve the usual statistical
\cite{stat} or causal (De Broglie - Bohm) \cite{holland}
interpretation of quantum mechanics, so that they are independent of
it. Nevertheless, they are very useful especially in the second frame;
in this section we modify the usual assumptions of the causal theory
to take into account the obtained results.

The starting point of the causal interpretation of quantum mechanics
can be formulated as follows: if we need a complex wave function to
describe a physical system, then it is reasonable to suppose that the
wave function itself and not just its modulus (i.e. also its phase or
quantities derived from this) has a direct physical meaning. In this
view, the statistical interpretation of $\psi$ (which is necessary,
since it is in accord with the experimental facts) is not necessarily
the only property which $\psi$ itself carries; the wave function can
indeed have a more potent role in the dynamics of a quantum system
\footnote{This idea was also present, in our opinion, in E. Majorana,
who considered electromagnetism on the same ground of the Dirac theory
of spin 1/2 particles (see ref. \cite{majo}). In this case, the
``Maxwell'' wave function is built up with the electric and magnetic
fields, which are direct physical quantities. The statistical
interpretation of $|\psi|^2$ is also recovered \cite{majo}.}.\\
The fundamental assumptions of the causal quantum theory of motion are
the following (see \cite{holland} for a general discussion):
\begin{itemize}
\item A physical system is described by a physical complex-valued
field $\psi (\x , t) \, = \, R \, e^{i S}$ which is
solution of the Schr\"{o}dinger equation;
\item $|\psi(\x , t)|^2 \, d^3 \x$ is the probability that a particle
described by the field $\psi ( \x ,t)$  lies between the points $\x$ and
$\x \, + \, d \x$ at time $t$;
\item The velocity of the particle is given by
\be
\bvec{v} (\x ,t) \; = \; \frac{1}{m} \, \gr S( \x , t)  \nonumber
\ee
The particle motion is univocally determined from the equation
$\frac{d \x}{d t} \, = \, \bvec{v}$ once the initial conditions
\bea
\x (0) & = & \x_0  \nonumber \\
\psi(\x , 0) & = & \psi_0(\x)  \nonumber
\eea
are given. The initial velocity of the particle is $\frac{d \x}{d
t}|_{t = 0} \, = \, \frac{1}{m} \, \gr S_0(\x) \, |_{\x = \x_0}$.
\end{itemize}
In the light of the results obtained here, it seems simple and
straightforward to generalize the previous assumptions by changing the
third postulate as follows:
\begin{itemize}
\item The absolute velocity of the particle is given by
\be
\bvec{v} \; = \; \bvec{v}_{||}
\; + \; \bvec{v}_{\perp} \; = \; \vb \; + \; \vs \, {\times} \, \s
\nonumber
\ee
where the ``drift velocity'' is
\be
\vb (\x ,t) \; = \; \frac{1}{m} \, \gr S( \x , t)  \nonumber
\ee
while the ``relative velocity'' is $\vs {\times} \s$ with
\be
\vs (\x ,t) \; = \; \frac{1}{2m} \, \, \frac{1}{R^2(\x , t)} \,
\gr R^2( \x , t)  \nonumber
\ee
and $\s$ is the spin direction of the particle (internal angular
momentum).\\
The total motion of the particle is made of an ``external'' motion,
given by $\frac{d \x_e}{d t} \, = \, \bvec{v}_B$, and an ``internal''
one, given by $\frac{d \x_i}{d t} \, = \, \vs {\times} \s$. It is
univocally determined once the initial conditions
\bea
\x_{tot} (0) & = & \x_0  \nonumber \\
\psi(\x , 0) & = & \psi_0(\x)  \nonumber
\eea
are given. The initial (total) velocity of the particle is
\be
\frac{d \x}{d t}|_{t = 0} \, = \, \frac{1}{m} \, \left(
\gr S_0(\x) \, + \, \frac{1}{2 R_0^2(\x)} \gr R_0^2(\x) {\times} \s
\right)_{\x = \x_0}  \nonumber  .
\ee
\end{itemize}
Let us observe that in the ``old'' formulation the knowledge of
$R(\x,0) \, = \, R_0(\x)$ determined only the initial probability
distribution, while here it directly determines also the initial
velocity.

Now, the point is: in which way the reformulation of the third
postulate is realized? Let us briefly discuss how the imposition of
the initial conditions determines the motion of the particle,\\
Solving the Schr\"{o}dinger equation and imposing the initial condition
$\psi(\x , 0) \, = \, \psi_0(\x)$, the functions $S(\x,t)$ and $R(\x,t)$
are exactly determined. From these, the fields $\vb(\x,t)$, $\vs(\x,t)$
and then $\bvec{v}(\x,t)$ can be built up. Then imposing also the
condition $\x(0) \, = \, \x_0$, one can exactly know the dynamical
trajectory $\x(t)$ from the equation
\be
\x(t) \; = \; \x_0 \; + \; \int_0^t \, \bvec{v} (\x(t'),t') \, d t'
\label{61}
\ee
Once $\x(t)$ is exactly know, also the velocities
$\vb(t) \, = \, \vb(\x(t),t)$, $\vs(t) \, = \, \vs(\x(t),t)$, and then
$\bvec{v}(t)$ are univocally determined. Instead, the dynamical
trajectories of the external and internal motion, given by
\be
\x_e(t) \; = \; \x_e(0) \; + \; \int_0^t \, \vb (\x(t'),t') \, d t'
\label{62}
\ee
\be
\x_i(t) \; = \; \x_i(0) \; + \; \int_0^t \, \bvec{v_{\perp}}
(\x(t'),t') \, d t' \label{63}
\ee
cannot be exactly known, since from the only initial condition
\be
\x(0) \; = \; \x_e(0) \; + \; \x_i(0)
\label{64}
\ee
it is not possible to deduce $\x_e(0)$, $\x_i(0)$. Then, in the
present reformulation the external and internal motions are not
exactly determined (we can obtain a family of possible motions, but
not the actual motion): only the complete motion can be univocally
determined.

But a question arises: what is the initial condition that one can
deduce from experiments? Is it really possible to assign $\x(0) \, =
\, \x_0$ in (\ref{64})?

In the ``old'' causal formulation, the internal motion was not
considered and the description of the (total) motion was given in
terms of an effective potential; the initial condition was imposed on
$\x_e$, coinciding (there) with $\x$.\\
It is reasonable to assume that the initial condition can be assigned
(for several experimental accidents) only on the external motion
(which is, in some sense) the mean motion of the particle), but not on
the complete motion. It is therefore interesting to see what happens
by keeping this point of view.\\
A direct consequence of this assumption, and of the fact that $S$,$R$
depend on $\x$ but not on $\x_e$ (as well as $\vb$, $\vs$ and $\bvec{v}$
do), is that the dynamical trajectory $\x(t)$ is no longer exactly
known. Consequently, also $\x_e(t)$ (besides $\x_i(t)$) is not
univocally determined by (\ref{62}) {\it even if}  $\x_e(0)$ is known.
This is a general result: if it is not possible to assign the initial
condition on $\x(t)$, then the motion (also only the external motion)
is no more univocally determined. We can only obtain a family of
possible motions, but not the real motion. In this framework, the
probabilistic interpretation of quantum mechanics (second postulate)
has therefore a ``natural'' assessment: the theory is intrinsically
deterministic, but our impossibility on fixing the initial condition
makes necessary a probabilistic formulation of it.

\section{Concluding remarks}

In this paper we have generalized the result obtained by Recami and Salesi
in ref. \cite{rs} and discussed its implications. We have shown that the
quantum effects present in the Schr\"{o}dinger equation are due to the presence
of a peculiar spatial direction associated with the particle that, assuming
the isotropy of space, we identify with the spin of the particle itself.
This result has been obtained by studying the invariance properties of the
equations of motion (especially the continuity equation). The picture
emerged from this analysis is the following: the motion of a quantum
particle is made of an ``external'' motion, described by the velocity field
$\vb$ defined in (\ref{26}), and an ``internal'' one (featured by the
presence of spin), driven by $\vs {\times} \s$, where $\vs$ is given in (\ref{27})
and $\s$ is the spin direction of the particle. In this framework, the so
called ``quantum potential'' is completely determined from the kinetic
energy of the internal motion. This allows to give a very simple and
natural explanations of some ``unusual'' properties attributed to the
quantum potential, as discussed in section 5.

The use of the quantum potential is particularly useful for the causal (De
Broglie - Bohm) interpretation of quantum mechanics; we have then
reformulated this interpretation to take into account the results obtained
in the previous sections, pointing out how this reformulation can be
realized in Nature. In particular, we discussed the physical implications
of the insurgence of an internal motion on the definition of the initial
conditions, which can be specified only on the external (mean) motion. This
{\it requires} necessarily a probabilistic formulation of quantum
mechanics. In this view, the internal motion is responsible both of the
quantum effects (described, properly, by the Hamilton-Jacobi equation
(\ref{33})) and of the quantum probabilistic interpretation of them (which
is allowed by the continuity equation (\ref{37}) and, in the present
formulation, by the fact that the internal motion has no effect on it
through the probability current density $\bvec{J}$ in (\ref{317})). This
result is, in some sense, a bridge between the causal and statistical
formulation of quantum mechanics: the theory is in its own deterministic,
but our impossibility to define proper initial conditions (caused by the
presence of an internal motion) forces to give a probabilistic
interpretation of it. The experimental confirmation of this probabilistic
formulation is then interpreted as an evidence of the internal motion for
quantum systems.

\vspace{1truecm}

\noindent
{\Large \bf Acknowledgements}\\
\noindent
The author is very grateful to Prof. Erasmo Recami and Dr. Giovanni
Salesi for useful discussions and their friendship.

\end{document}